# Pore morphology evolution and atom distribution of doped $Fe_2O_3$ foams developed by freeze-casting after redox cycling

P.J. LLoreda-Jurado, Jesús Hernández-Saz, E. Chicardi, A. Paúl, R. Sepúlveda*.

Departamento de Ingeniería y Ciencia de los Materiales y del Transporte, E.T.S. de Ingenieros, Universidad de Sevilla, Avda. Camino de los Descubrimientos s/n., 41092 Sevilla, Spain.

*Corresponding author: rsepulveda@us.es

Abstract

Chemical looping water splitting systems operate at relatively high temperatures (450-800 ºC) to produce, purify, or store hydrogen by the cyclic reduction and oxidation (redox) of a solid oxygen carrier. Therefore, to improve long-term operation, it is necessary to develop highly stable oxygen carriers with large specific surface areas. In this work, highly interconnected doped $Fe_2O_3$ foams are fabricated through the freeze-casting technique, and the aid of a submicrometric camphene-based suspension to prevent Fe sintering and pore clogging during redox operation. The influence of the dopant elements (Al and Ce) over the pore morphology evolution, and redox performances are examined. The use of an $Fe_2O_3$ porous structure with initial pore size above 100 microns shows a significant reduction of the sample densification, and the addition of $Al_2O_3$ by the co-precipitation process proves to be beneficial in preventing the generation of a core-shell structure following redox processing.

Keywords: Chemical looping system, $Fe_2O_3$, pore evolution, freeze-casting, camphene, $Al_2O_3$.

1. Introduction

In recent decades, the reduction of $CO_2$ emissions from fossil-fuel consumption has become a necessity due to the undeniable effects of global warming [1]. Switching to a clean and renewable energy source is an unavoidable route to counterbalance this harmful effect. Hydrogen seems a valid alternative, since no $CO_2$ is emitted at its consumption, although current large-scale production is still based on reforming methane, oil, or coal [2]. Therefore, a more realistic option for cleaner hydrogen production involves utilising renewable (or $CO_2$ neutral) resources in combination with carbon-capture systems [3]. The Chemical-Looping Water-Splitting process (CLWS) or Steam-Iron Process (SIP), wherein iron is employed, are promising technologies for hydrogen production and/or purification that can combine both these technological requirements. The SIP is a two-step process, where first a suitable solid oxide (i.e., an oxygen carrier) is exposed to a reducing gas stream to turn it into metal iron. The gas stream could be a mixture of hydrogen and CO, commonly referred to as syngas [4]. A steam flow then re-oxidizes the Fe into iron oxide and produces pure hydrogen by splitting the water molecule. Finally, the re-oxidized iron oxide is employed to restart the process.

Hematite ($Fe_2O_3$) presents an attractive oxygen carrier for the production, purification, or storage of hydrogen by means of the SIP since it is abundant and inexpensive and also due to its redox properties [5–7]. Iron could act as a secondary source of hydrogen storage if it is oxidized by steam to form $Fe_3O_4$ and if the hydrogen released is then collected, theoretically, at standard temperature and pressure conditions, whereby 0.534 L of $H_2$ can be stored per gram, which is comparable with the storage density of conventional metal hydrides [8]. Nonetheless, the performance of $Fe_2O_3$ as an oxygen carrier is deeply affected by the sintering phenomena, also called deactivation, since many applications require 2numerous redox cycles at high temperatures.

In order to prevent or delay particle sintering and to control the redox reaction kinetic, many studies have doped the hematite with diverse elements, such as Al [9], Ce [10, 11], Zr [12], Mo [13], and Ca [14]. However, these dopant elements are insoluble (Ce, Zr) or only partially soluble (Al, Mo, Ca) in Fe. The thermal cycles of the $Fe_2O_3$ would lead to a major depletion in metal atoms, which is especially drastic with Ce and Al. Other researchers have demonstrated the use of structural porous materials as oxygen carriers to replace the fluidized-bed systems. Honeycomb monoliths [10, 15] and $Fe_2O_3$ foams [16] proved to possess comparable redox performances.

$CeO_2$ as an additive induces crystal modification and stabilisation of the iron oxides [17], despite the limited solubility observed among $Fe_xO$ and $CeO_2$ phases. A solid solution has been obtained, since $Fe^{3+}$ ions can be incorporated at the $CeO_2$ fluorite crystal structure at lattice ($Ce^{4+}$) and interstitial position by hydrothermal [18] or co-precipitation [19] routes rather than by the standard milling process. Furthermore, during high temperature operation, the formation of hematite-like solid solutions and perovskite $CeFeO_3$ phases ($Ce^{3+}$) is promoted. Therefore, the increase in vacancy concentration enhances the redox performance of $Fe_2O_3$-$CeO_2$ systems by higher oxygen mobility [20]. $Al_2O_3$, as an additive, improves the cycle stability of $Fe_2O_3$ during continuous operation [7]. $Fe_2O_3$-$Al_2O_3$ shows limit solubility: at 1300 ºC, about 20 wt.% $Al_2O_3$ can be incorporated into a hematite-like solid solution ($H_{ss}$) [21], while at room temperature, solubility is neglected. The thermal cycling induced the formation of hercynite ($FeAl_2O_4$) [22], which has a spinel crystal structure 1where Fe and Al occupy divalent and trivalent cation positions. This spinel phase is more favourably formed during the reduction reaction and early stage of the oxidation reaction since Fe and FeO phases react exothermically in the presence of $Al_2O_3$. In the case of the $Fe_3O_4$-$Al_2O_3$ system, a miscibility gap is formed below 860 ºC and two phases are found: magnetite-based solid solution ($M_{ss}$) and hercynite. At 750 ºC, $M_{ss}$ shows a solubility maximum of 12 wt.%. The use of up to 3 wt% $Al_2O_3$, enhances the reduction process of $Fe_3O_4$ due to the formation of a network-like structure of wüstite [23, 24].

Freeze-casting (FC) is a promising manufacturing technique for the creation of porous materials with a wide variety of porosity range, pore size, and morphology [25]. It is also cost-effective, eco-friendly, and scalable on an industrial level. However, pure Fe-foams, manufactured by FC using water [26] and camphene-based [27, 28] hematite slurries, have shown significant modification of the pore structure after being subjected to several redox cycles. Authors have found a continuous reduction in the pore size due to the sintering of the Fe cell walls, which is especially critical when initial pore sizes are under 30 μm in diameter. Nevertheless, Lloreda-Jurado *et al.* [29] have reported doped $Fe_2O_3$ foams manufactured by FC with diameter pore sizes from 30 to 130 μm, simply by controlling the processing parameters.

Therefore, the aim of this research is to determine the influence of the redox cycling over the pore morphology and atom distribution on doped $Fe_2O_3$ foams manufactured by FC with initial pore sizes above 100 μm. $Fe_2O_3$ powder doped with Al and Ce was synthesized by the co-precipitation route and subsequently suspended in camphene.

2. Experimental Procedure

2.1. Powder synthesis

Doped $Fe_2O_3$ powder is fabricated using the co-precipitation route previously described by Lorente *et al.* [30]. Initially, aqueous solutions of selected metal nitrates and citric acid are stirred at 80 ºC until a gel is obtained. The resulting gel is dried at 60 ºC overnight and then calcined at 350 ºC for 2 h followed by a final step of 800 ºC for 8 h to ensure a good level of crystallinity. The resulting powder, *as-synthesized,* is later crushed and sieved to separate particles within 160-200 µm in diameter. The final powder composition is approximately 98 wt% $Fe_2O_3$, 1.75 wt% $Al_2O_3$, and 0.25 wt% $CeO_2$.

2.2. Sample preparation

Doped $Fe_2O_3$ foams are fabricated using the freeze-casting method previously described by Lloreda-Jurado *et al.* [29]. The method starts by a further reduction of the particle size by ball milling down to 0.2 µm in diameter. A 5 vol.% powder suspension in camphene (Sigma Aldrich, Spain) is then obtained by ball-milling at 60 ºC for approximately 12 h. First 3 wt% of dispersant agent KD4 (oligomeric polyester provided by CRODA Ibérica) is mixed for 30 min in camphene, and the powder is subsequently incorporated and ball-milled for 8 h. Finally, 20 vol.% polystyrene (Sigma Aldrich, Spain) with a $M_W$ = 350,000 g·$mol^{-1}$ was added as a binder and mixed for another 3 h. Both organic additives are formulated according to the initial powder load and are incorporated sequentially to aid in the proper particle dispersion. The camphene suspension was then poured into a mould of 30 mm in diameter by 15 mm in height PFTE, preheated at 60 ºC placed inside an incubator. Directional solidification was promoted by running water at 42.5 ºC through the mould base and gradually reducing the incubator temperature until reaching complete sample solidification. A slow solidification rate was employed to enhance pore enlargement during the FC process. Finally, after demoulding, and a 3-day camphene sublimation process at ambient conditions under forced airflow, the sample is sintered in air at 600 °C for 2 h for organic-specimen burn-out, plus a sintering of 1100 °C for 2 h for particle sintering. Subsequent to the sintering, the sample *as-fabricated measures* approximately 9 mm in height by 24 mm in diameter.

2.3. Microstructural and redox performance characterisation.

The pore morphology characterisation of the as-fabricated sample is determined by Optical Microscopy (OM) and Field Emission Scanning Electron Microscopy (FESEM) across the entire height. Initial pore size and cell wall width are determined by OM images using the non-redundant maximum-sphere-fitted algorithm [31] of the ImageJ software.

In turn, the redox behaviour is determined on subsamples of 1.2 x 1.2 x 4.5 mm in volume sectioned from the centre-top part of the as-fabricated sample (to reduce the presence of pores under 50 µm in diameter), while maintaining the freezing direction parallel to the longest dimension. Some additional manual grinding is necessary to obtain a 20 mg subsample. This study is carried out in a thermogravimetric analyser (TG) Netzsch® STA 449 F3 Jupiter using an alumina plate as support for the porous subsample. Two redox conditions, dynamic and isothermal, are analysed in accordance with the following setup: 1) Dynamic: reduction up to 700 °C with a gas flow of $N_2$-20 vol.% $H_2$, followed by an oxidation up to 800 °C with a gas flow of $N_2$-12 vol.% $H_2O$; 2) Isothermal: 3 consecutive cycles of reduction at 750 °C, and oxidation at 450 °C. Heating rate and flowrates are established at 5 °C/min and 100 NmL/min, respectively. In both redox conditions, the porous samples ae removed from the STA after the oxidation stage is completed.

Subsequent to the redox test, samples are subjected to X-ray computed tomography (X-CT) in order to determine the pore morphology variation. Optical configuration achieves a voxel of 7.56 µm/pixel, the collected 2D cross-sectional images are pre-processed with ImageJ software, and 3D visualisations are made using Avizo software.

Dopant element distribution is determined by Transmission Electron Microscopy (TEM) in the as-synthesised and as-fabricated condition. An FEI TALOS F200S is used operating at 200 kV with a field emission gun and equipped with Super-X dispersive X-ray spectrometry systems including two silicon drift detectors (FEI ChemiSTEM), provided with their specific software for acquisition and data processing (Thermo Scientific Velox®). The Crystallography Open Database is utilised to identify the unknown planes of the $CeO_2$ and iron oxide particles.

Subsequently, X-ray Photoelectron Spectroscopy (XPS) is employed to determine the oxidation state and chemical contents of doped $Fe_2O_3$ in the as-synthesised, as-fabricated, and *after isothermal redox processing*. A SPECS Hemispherical Energy Analyzer model Phoibos 150 MCD is used. The deconvolution of XPS spectra is performed in accordance with the Gaussian-Lorentzian Cross Product expression as a fitting function (E.q.1) using PeakFit software (Systat Software Inc.)

$$y = \frac{a_0}{1+a_3\left(\frac{x-a_1}{a_2}\right)^2 exp\left[(1-a_3)\frac{1}{2}\left(\frac{x-a_1}{a_2}\right)^2\right]} \qquad (1)$$

where $a_0$ is the amplitude, $a_1$ is the centre of the curve, $a_2$ is the width ($a_2 >0$), and $a_3$ is the shape of the curve ($0 \leq a_3 \leq 1$, the pure Lorentzian occurs with $a_3 =1$ and the pure Gaussian occurs with $a_3 =0$ )

Finally, to corroborate, phase evolution in the as-fabricated and after isothermal redox processing is determined by X-Ray Diffraction (XRD). Previous to the XRD study carried out, the specimens are mechanically milled to ensure the sample representativeness. XRD patterns are recorded using a PANalytical X'Pert Pro instrument (Malvern Panalytical Ltd, Malvern, UK) equipped with a Bragg-Brentano θ/θ geometry, a Cu Kα radiation source (45 kV, 40 mA), a secondary Kβ filter and an X'Celerator detector. The XRD patterns are obtained by scanning between 20 and 120º of the 2theta degree with 0.03º steps and a counting time of 800 s·step-1. The open-access Crystallography Open Database (COD) is employed to elucidate the phases developed with the corresponding structures and space-group symmetries (SGSs). A lanthanum hexaboride, LaB$_6$ (Standard Reference Material 660b, NIST) pattern is utilised to calibrate the positions of the diffraction lines.

3. Results and discussion

3.1 Doped Fe$_2$O$_3$ powder characterisation: *as-synthesised*

Figure 1a shows a High-Angle Annular Dark-Field (HAADF) image in Scanning TEM mode obtained from an aggregate of doped Fe$_2$O$_3$ particles that exhibit an average size of 0.2 μm in diameter. The element distribution analysis performed by EDX shows a uniform distribution of Fe, O, and Al (Fig. 1b-c), while Ce is segregated to a different location at the particle surface (Fig. 1e). These results suggest that Ce atoms are depleted from the Fe$_2$O$_3$ matrix during particle synthesis. This is reasonable since oxide phases can be separated during annealing at temperatures above 600 ºC [32, 33]. The doped Fe$_2$O$_3$ particle (Fig. 1f) shows an interplanar distance of 0.251 nm (Fig. 1g), in agreement with the plane 110 of pure bulk $\alpha$-Fe$_2$O$_3$. Some minor lattice distortion is detected in other interplanar distances. This is due to Al ions (with smaller ionic radii) randomly occupying cation positions in the rhombohedral crystal structure of $\alpha$-Fe$_2$O$_3$. This result corroborates the formation of a (Fe,Al)$_2$O$_3$ solid solution described in the literature [21, 34].

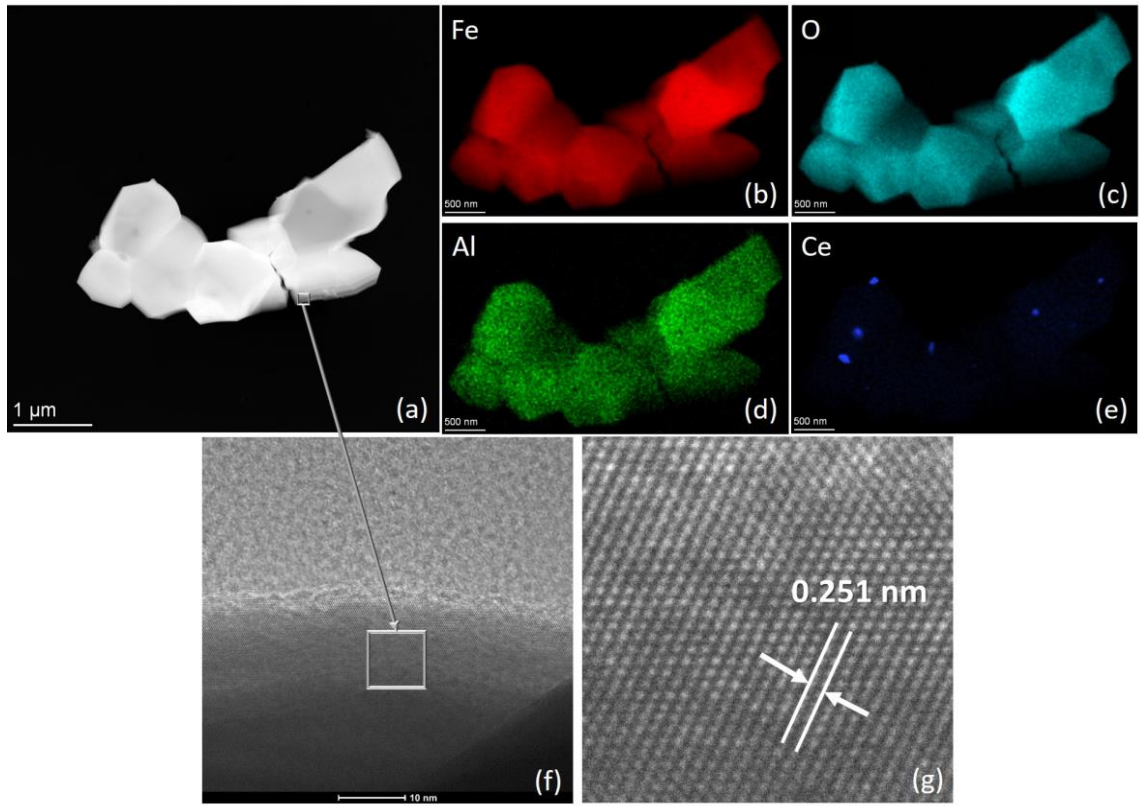

Figure 1. HAADF-STEM image of the initial doped $Fe_2O_3$ powder (*as-synthesised*) showing an aggregate of particles. (a) EDX analysis shows the distribution of the principal elements: Fe (b), O (c), Al (d), and Ce (e). An HRTEM image of a selected area of the particle (f), displays the atomic arrangement of the crystal plane (110) (g).

3.2. Doped $Fe_2O_3$ foam characterisation: *as-fabricated*.

During FC, the directional solidification imposed on the slurry created a unique pore structure. The growing camphene dendrites push the particles into the interdendritic spaces and form particle aggregates that replicate the dendrite morphology. Since the sample heat is extracted through the bottom by a constant-temperature flow of water during solidification, the incubator temperature is also constantly reduced. The thermal gradient decreases and the solidification front velocity increases across the sample height. This situation promotes the formation of an equiaxed and interconnected porous structure, where the average pore size drastically increases with the sample height as the solidification front moves a few millimetres away from the sample base, as shown in Figure 2a-b. Average pore sizes are calculated in areas described by Figure 2a-d, as 29 ± 12, 100 ±39, 117 ± 44, and 127 ± 43 µm, respectively. The average thickness of the cell wall shows a much wider size distribution along the sample, from minimum values of 3-4 µm to a maximum of 60 µm.

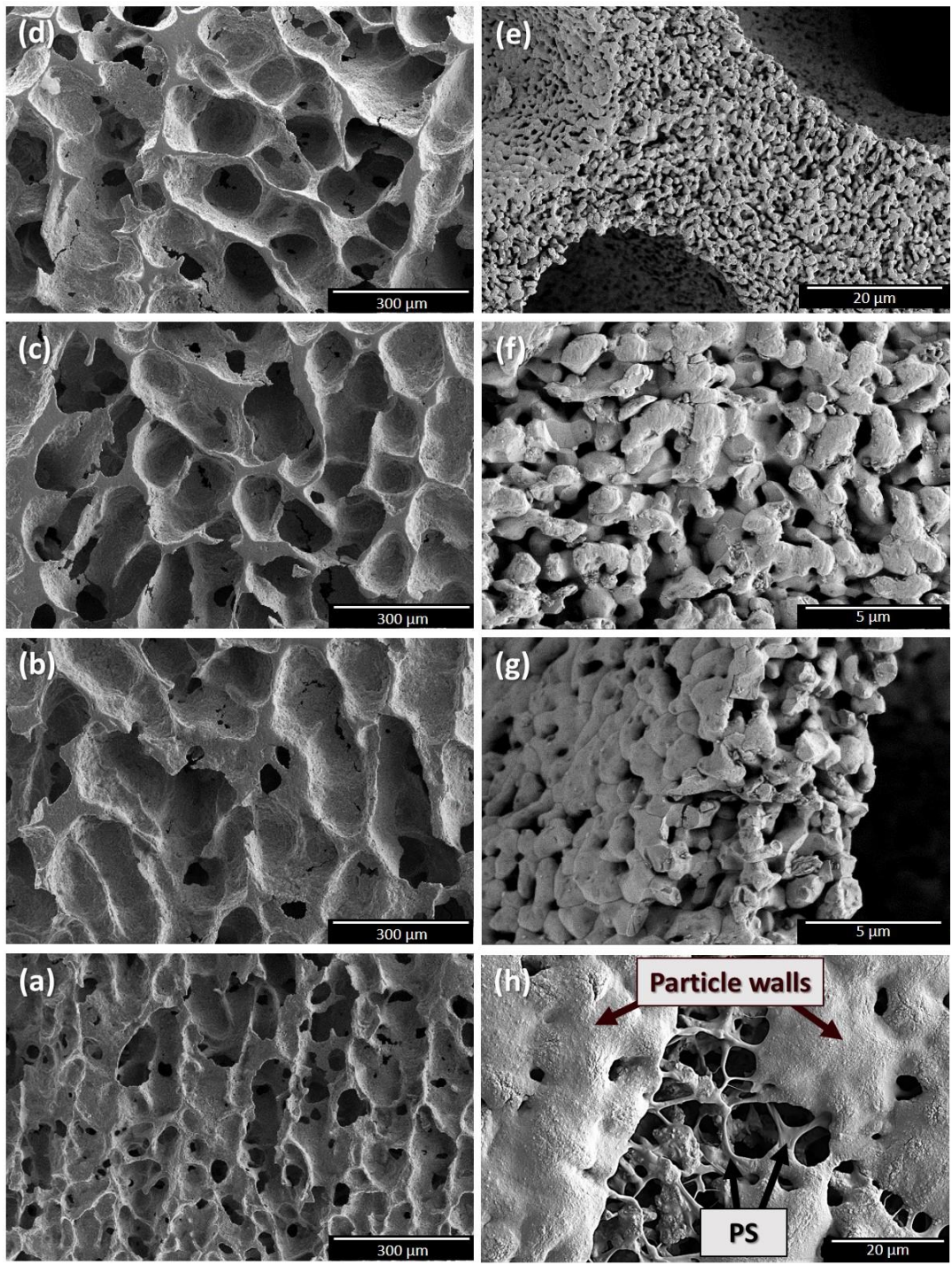

Figure 2. SEM micrographs of the Doped $Fe_2O_3$ foams (*as-fabricated*) at different sample heights: 0 (a), 3 (b), 6 (c), and 9 mm (d). Details of the porous cell-wall surface (e,f) and cross-section are shown (g). Micrographs a-d are taken using the same magnification. Doped $Fe_2O_3$ foams before sintering (h), particles, and some PS become aggregated into the walls, while a significant PS forms struts across the walls.

Regardless of the position, the cell wall (Fig. 2e) exhibits a highly porous structure on its own. An interconnected pore structure with diameters less than 1 µm is observed at the cell-wall surface (Fig. 2e-f) and its cross-section (Fig. 2g). This secondary porosity is attributable to the pyrolization of C-based compound incorporated through the foam fabrication. During slurry preparation, dispersant agent and PS are added at different moments of the ball-milling procedure. First, the dispersant agent is added close to the beginning of the dispersion process, and after 8 h of ball-milling the polystyrene (PS) is incorporated. KD4 is an oligomeric polyester with a C66 long chain, and displays good adsorption properties due to its carboxylic acid group. This carboxylic group is anchored to the doped $Fe_2O_3$ particle surfaces leading to a high steric barrier approximately above 10 nm [35]. Since PS has phenyl functional groups every two C atoms and camphene is a bicyclic skeleton monoterpene, both dispersant and binder show good chemical affinity. The melted camphene forms a non-polar solvent where the PS polymer chains are disentangled (dissolved). Doped $Fe_2O_3$ particles could be trapped by the PS long C-C chain due to the hydrogen bond driven by the KD4 previously anchored. Moreover, during directional solidification of camphene, 20 vol.% PS was incorporated into the particles walls and formed struts across them, as showed in Figure 2h.

The sintering process volatilises the C-based species and some carbonaceous reductant gases are produced at even relatively low temperatures [36]. KD4 and PS have a chemical formulation of $C_{90}H_{172}O_{10}$ and $(C_8H_8)_{n-1}$ respectively, and therefore a certain level of carbothermic reduction could be expected according to Bell's Diagram for the Fe–C–O [37], where the magnetite phase remains stable at temperatures lower than 570ºC and $CO/(CO+CO_2)$ gas ratio below 0.5, but no traces of $Fe_3O_4$ is detected by TEM analysis subsequent to sintering (i.e., a nearly 0% degree of reduction). In this work, the mole ratio of C/O described as the carbon contend at the KD4 and PS by the oxygen at the hematite is equal to 0.2, which is lower than the value referred to in other studies (0.63 [36] or 1 [38]) with a degree of reduction above 55%. This result suggests that reductant gases produced during the sintering process, even the close integration between solid particle and C-based species achieved by the slurry preparation, are insufficient to promote a significant reduction reaction, despite the close integration between solid particle and C-based species during slurry preparation and the FC procedure.

Figure 3a shows an HAADF-STEM image of an aggregate of particles where the sintering exerts no significant effects on the doped element distribution. The EDX analysis performed (Fig. 3b-e) on several particles shows a persistent homogeneous distribution of the Al atoms (Fig. 3d), and Ce atoms forming individual clusters (Fig. 3e). The HRTEM image of Figure 3f shows the crystalline structure of both particles. A detailed magnification of the $CeO_2$ particles is shown in Figure 3g. The distant measures correspond to planes (111) of stoichiometric $CeO_2$.

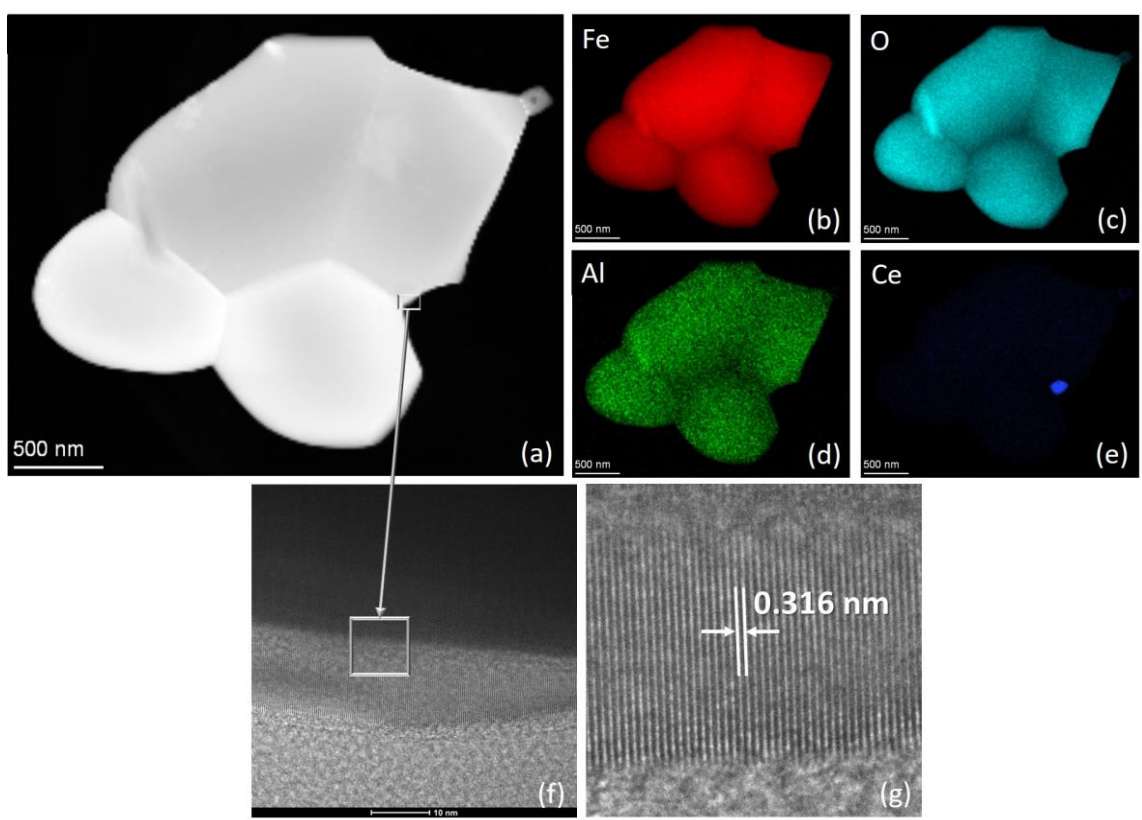

Figure 3. HAADF-STEM image of the doped $Fe_2O_3$ foam (*as-fabricated*) showing an aggregate of particles (a). EDX analysis shows the principal element distribution: Fe (b), O (c), Al (d) and Ce (e). The HRTEM image of the selected area of a $CeO_2$ particle on (f) displays the atomic arrangement of the crystal plane (111) (g)

3.3. Doped $Fe_2O_3$ foam characterisation, *subsequent to the redox process.*

During dynamic redox processing (Figure 4a), doped $Fe_2O_3$ foam initiates its reduction at 448 ºC, which is a higher temperature than that of the powder version (300 ºC), and the step reaction of hematite to magnetite to iron is inhibited. However, both were completed around 600 ºC. At first glance, this result shows the effect of the sample morphology, the oxygen release has been delayed to a higher temperature, and the kinetic reaction is increased. While the temperature kept rising from 600 ºC to 700 ºC, a sintering process on the formed Fe walls is definitely encouraged. Since most redox processing under SIP conditions are performed at temperatures ranking from 700 ºC to 900 ºC, there is a clear evidence that regardless of the morphology of the oxygen carrier, a sintering process (i.e., efficiency loss) is inevitably encountered. Through the oxidation reaction, both foam and powder reach a maximum weight gain corresponding to $Fe_3O_4$: this oxide phase is already predicted by the Bell Diagram [37], and has previously been reported by other authors [37, 39] in the case of steam oxidation. The use of a foam morphology shows a decrease on the kinetic reaction, while the subsequent formation of iron oxides layers on top of the Fe cell walls limit the access of steam to fresh Fe. Moreover, at 634 ºC, the oxidation reaction to $Fe_3O_4$ could be considered completed.

The isothermal redox processing (Figure 4b) shows weight loss and gain with the alternate reduction and oxidation reaction in the doped foam and powder, respectively. As discussed previously, the oxidation reaction at 750 ºC converts $Fe_2O_3$ into metallic Fe, while the steam re-oxidation at 450 ºC allows the conversion to $Fe_3O_4$. Thus, a maximum theoretical value of 97% weight change could be achievable during oxide conversion. In the first redox cycle, $Fe_2O_3$ turns suddenly into Fe in just 5 min within the initial 30 min of reduction. However, the oxidation step takes almost 1 h to recover 88% of the weight lost following a power-law of $\sim 3.46 \cdot t^{0.5}$ ($R^2$=0.94, with $t$ the time in minutes). With the second and third redox cycle, the reduction reaction is again abrupt and nearly completed to metallic Fe in minutes, and the re-oxidation reactions takes place in 1 h, whereby 86% of the weight lost is recovered. In both cases, the re-oxidation reaction shows different power-law behaviours: roughly linear $\sim 0.2 \cdot t$ ($R^2$=0.97) for the second step, and $\sim 5.42 \cdot 10^{-3} \cdot t^{1.5}$ ($R^2$=0.98) for the third step. These results suggest that the re-oxidation reaction is accelerated in the redox cycles due to the enhanced access of the steam granted by the interconnected porous structure of the foam. In contrast with the powder version, where reduction and oxidation reaction follows the same trend as the foams, the redox performance dropped significantly with the cycling, from 92% in the first cycle to 85% in the third cycle. The doped $F_2O_3$ has demonstrated a more regular and steadier redox performance, by 1mitigating the decrease in the oxygen weight gain in the redox cycles. Additionally, the foam is entirely converse to metallic Fe after every reduction reaction, as is the re-oxidation reaction which is primarily limited by the access to fresh Fe, among other thermodynamic considerations. Indeed, the decrease in redox performance is usually attributed to a variety of factors, such as: 1) the 1reduction temperature, especially above 600 ºC, promotes the sintering of the Fe walls 1thereby reducing the porosity, especially those observed in Figure 2e-g; 2) The metallic Fe also creates a layer at the cell-wall surface [40], which segregates Ce and Al oxides species into the grain boundaries due to the lack of solubility and ultimately to the wall interiors . As discussed earlier, Ce atoms are segregated from the $(Fe,Al)_2O_3$ solid solution during the powder synthesis and foam consolidation heat treatments. Therefore, the improvement in the oxygen mobility and storage capacity expected from the addition of $CeO_2$ [41, 42] could decrease due to the reduction in the number of interfaces $Fe_3O_4/CeO_2$. Likewise, aluminium atoms are highly scattered in the doped $Fe_2O_3$ powder and foam due to the co-precipitation method employed during their synthesis. $Al_2O_3$ prevents the formation of a dense iron shell by promoting a reduction step in the sample surface, from magnetite to a network-like structure of wüstite, and, finally, to iron [24, 43], thereby delaying the sintering of the fresh Fe particles [44]. These aspects could counterbalance the noteworthy metal sintering during the reduction step, and decrease the loss of redox performance.

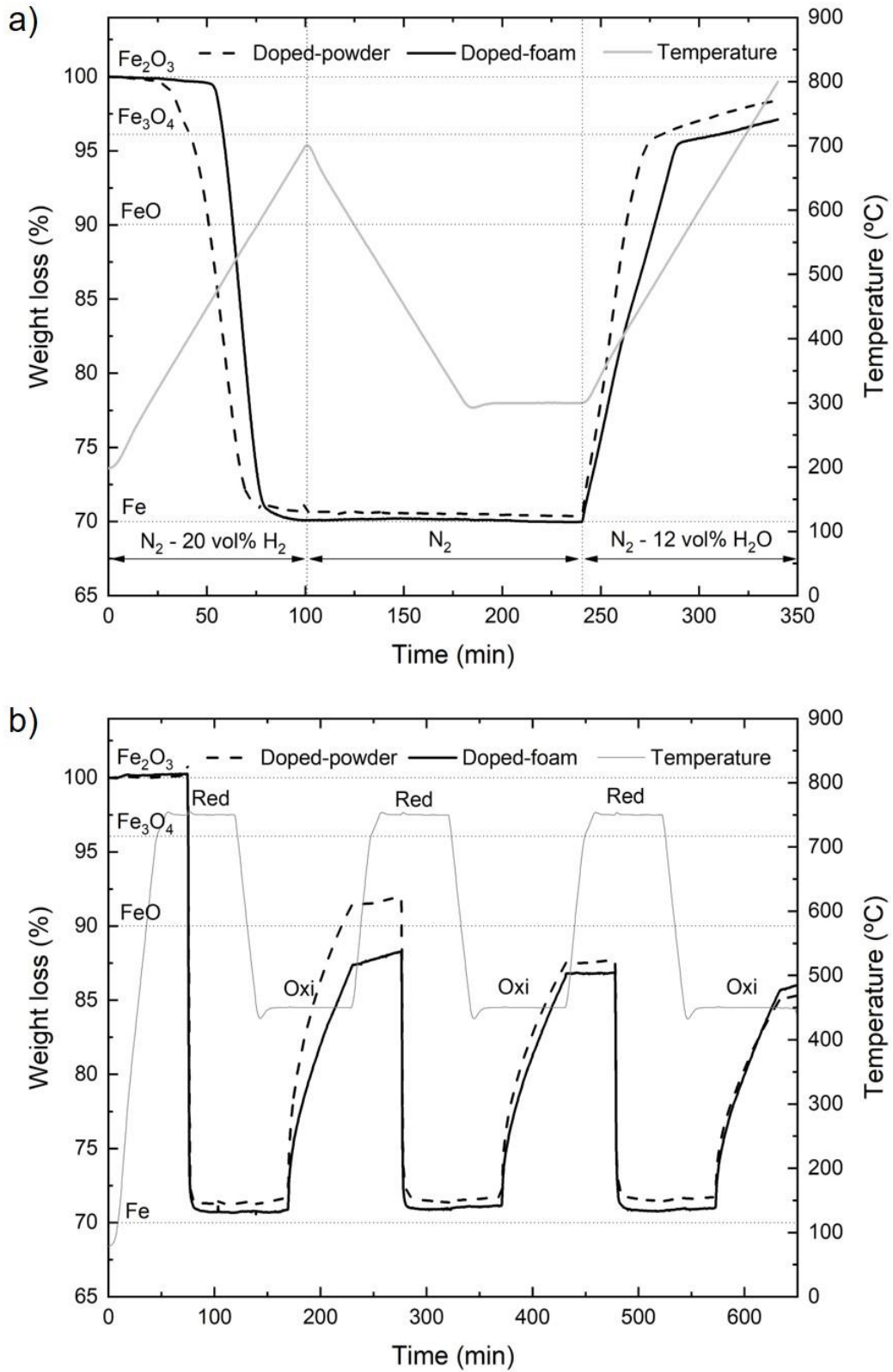

Figure 4. Weight changes of Doped $Fe_2O_3$ foam under dynamic (a) and isothermal (b) redox condition. The doped powder behaviour has been included as a comparison. $Fe_2O_3$, $Fe_3O_4$, FeO, and Fe theoretical weight lines are incorporated as reaction guidelines.

The evolution of the pore morphology of the samples subjected to dynamic and isothermal conditions are characterised by X-CT. The subsamples extracted from the doped $Fe_2O_3$ foams correspond to the top half of the entire sample, where the initial pore size varies from 110 μm at the base to 143 μm at the top. Figure 5a shows the 3D reconstruction of the complete doped $Fe_2O_3$ foams subsequent to the dynamic redox cycles. No significant shrinkage or crack formation is observed. During constant heating, the dynamic redox condition induced a complete reduction of the doped $Fe_2O_3$ foam to Fe, and a re-oxidation to $Fe_3O_4$ (magnetite) phase due to thermodynamic limitation. As the slice reconstructions (Fig. 5b-d) show, the pore morphology was kept constant across the sample height. The initial equiaxed and interconnected porous structure observed in Figure 2 was retained. Moreover, the pore size slightly varies as compared to the initial pore size: the 2D projections show an average pore size of 96, 137, 150, 149, 165, and 157 μm from the bottom to the top specimen, respectively. The average thickness of the cell wall is determined as 48 μm at the sample bottom, and decreases continuously until it reaches 30 μm at the top. This value is overestimated since no cell wall below 20 μm in thickness was detected by this measurement technique, due to the threshold of detection being a voxel size of 7.56 μm. The reduction of the thinner cell walls (>20 μm) might counterbalance the overall shrinkage and expansion effect during this redox cycle. The $Fe_2O_3$ thinner cell walls could be rapidly integrated into thicker walls at the reduction stage, via the sintering of Fe particles. These results show the minor effect of the dynamic redox processing on the doped $Fe_2O_3$ foam microstructure, and the beneficial effects of the interconnected pore structure developed by the freeze-casting technique employed in this research, where no significant sample densification or pore clogging are produced.

A distinguishing feature is observed in the sample, which is especially evident in the top-half zone: the formation of a cell wall aperture that creates a new close porosity. These effects have previously been reported [26] on pure Fe samples, but are described here for the first time for doped $Fe_2O_3$ foams. During reduction, the porosity of the cell walls improves the removal of oxygen, by improving the access to more oxidised material. At peak reduction temperature (700 ºC), a complete reduction to metal Fe is feasible and expected, the sintering of Fe particles would be undoubtedly promoted, and the cell walls become denser. In the oxidation stage, the constant formation of $Fe_xO$ phases produces a local volumetric expansion due to the crystal phase conversion and the diffusion of Fe atoms outwards, especially when the re-oxidation leads to the formation of an $Fe_2O_3$ phase. To accommodate these new phases, they are separated and, eventually, exfoliated from the remaining Fe base. This morphology feature is more evident in the top part of the dynamic redox samples (Figure 5c-d). As pore size increases, Fe walls can act as flat surfaces since their curvature ratio also increases. Therefore, the local expansion of the new phases could be released normal to the surface and then be exfoliated from the Fe base. When pore size or curvature ratio decreases, the Fe walls might resemble a cylinder, and the same volume expansion could no longer be released normal to the surface, and the $Fe_xO$ grains tend to sinter into a more compact layer, thereby diminishing any opportunity for exfoliation.

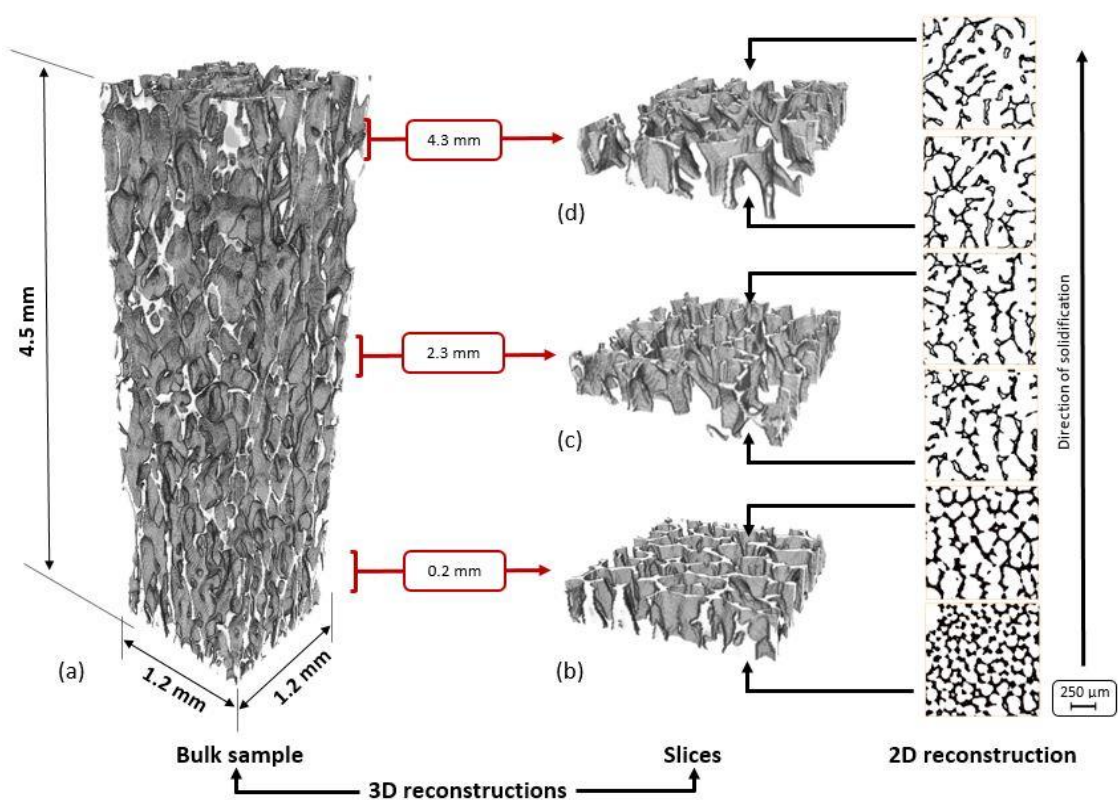

Figure 5. 3D reconstructions from X-CT of the doped $Fe_2O_3$ foams after dynamic redox processing: bulk sample (a) and slices of 250 µm in thickness at different sample heights: (b) 0.2 mm, (c) 2.3 mm and (d) 4.3 mm. The 2D projections are the bottom and top part of the slice reconstructions showing the pore-size evolution.

After 3 complete redox cycles under isothermal conditions, the doped $Fe_2O_3$ bulk sample retains the overall interconnected pore structure, as shown in Figure 6a. The 2D projections (Figure 6b-d) show average pore sizes of: 56±16, 76±21, 132±34, 138±36, 117±29, and 120±37 µm from the bottom to the top of the sample, respectively. The average thickness of the cell wall shows a higher value at the bottom (55 µm) and top (50 µm) parts, as compared with the centre (40 µm); as discussed previously, this value is overestimated due to the measurement technique. In this case, a significant reduction in pore size was observed at the bottom part of the sample, where the initial average pore size passes from nearly 100 µm to 56 µm. As the redox cycles occurred, during the reduction step, the Fe walls thickened, and the pore size and overall porosity decreased due to the rapid sintering at 750 °C. This situation is enhanced at the bottom of the sample, as can be seen in Figure 6b.

In contrast, the doped foam sample under the dynamic redox condition shows no cell wall aperture across the sample height, which suggests a decrease in the Fe outward mobility to form $Fe_xO$ due to the lower re-oxidation temperature (450 °C), which reduces the possibility of vacancy formation in the cell-wall interiors. In isothermal conditions, the reduction step leads to the formation of $Fe_3O_4$ preferentially, which decreases the volume expansion during the oxidation reaction and a possible exfoliation. Furthermore, the addition of $Al_2O_3$ in pure Fe has shown a reduction in the pore formation due to the Kirkendall effect with the conversion of Fe to magnetite [43]. The vacancy diffusion in magnetite is four times lower than in hematite [45], and therefore the Fe outward mobility is considerably reduced.

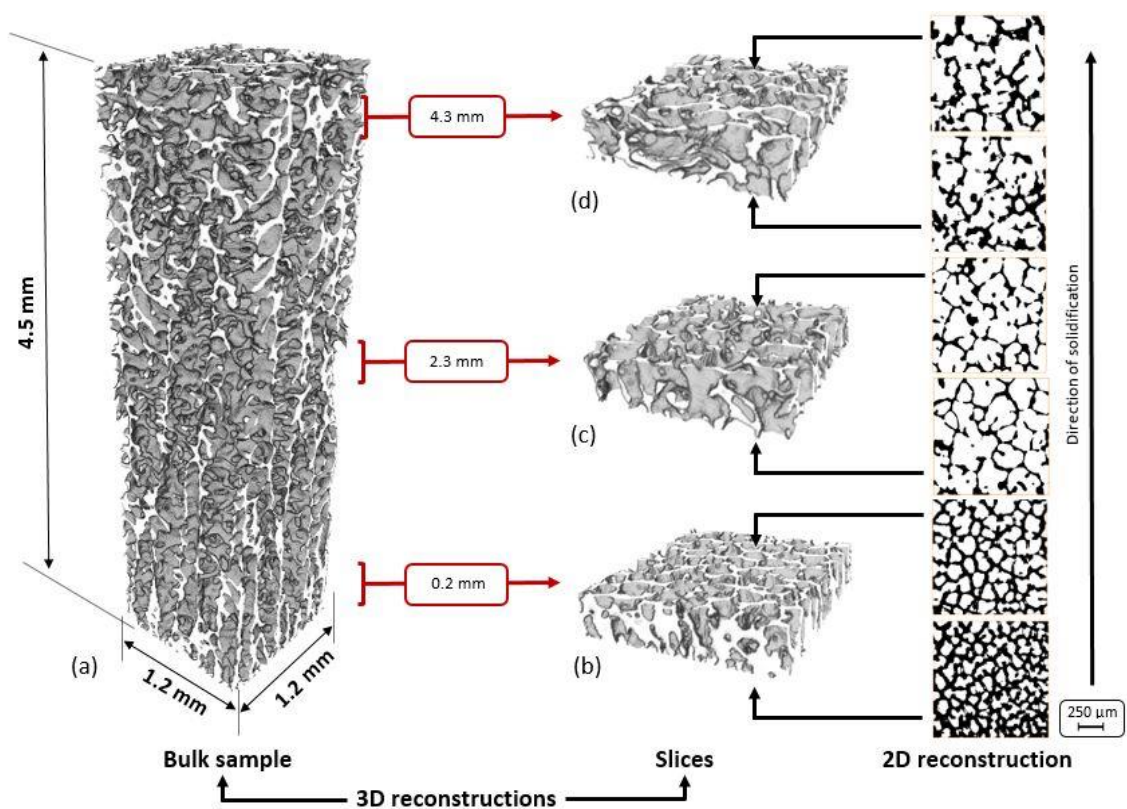

Figure 6. 3D reconstructions from X-CT of the doped $Fe_2O_3$ foams subsequent to isothermal redox processing: bulk sample (a) and slices of 250 µm in thickness at different sample heights: (b) 0.2 mm, (c) 2.3 mm, and (d) 4.3 mm. The 2D projections are the bottom and top part of the slice reconstructions showing the pore-size evolution.

In each re-oxidation step, some metallic Fe remains unconverted, as shown in Figure 4b. Therefore, assuming a re-oxidation reaction from Fe to $Fe_3O_4$, the remaining metal could be estimated in 9.3 wt% after the first cycle and 11.3 wt% for the remaining second and third cycles, which yield a 6.35 vol.% and 7.80 vol.%, respectively. Figure 7a shows a SEM micrograph of the doped $Fe_2O_3$ foam surface after the isothermal redox condition. The faceted surface of the magnetite grains is interrupted by a spider-like Fe network; these Fe phases correspond to the unconverted metal. An XRD analysis (Figure 7b) under both the *as-fabricated* and *after redox process* conditions is carried out in order to corroborate this assertion. This analysis detects only the hematite phase ($Fe_2O_3$, R-3c, ref.no. 1546383 in the COD) in the as-sintered specimen. In turn, for the *after redox process* two majority phases are elucidated, magnetite ($Fe_3O_4$, Fd-3m, ref. no. 7228110 in COD) and ferrite ($\alpha$-Fe, Im-3m, ref. no. 9016601 in the COD). Moreover, slight peaks of unreacted hematite are also detected in this specimen. All these observations therefore corroborate that the initial $Fe_2O_3$ evolves into $Fe_3O_4$ with some elemental Fe subsequent to the redox cycles. Furthermore, the position of the $Fe_2O_3$ and $Fe_3O_4$ peaks are not appreciably displaced, and show no influence of Al or Ce. However, the lack of a core-shell of metal does indeed corroborate the influence of the Al atom clusters well-dispersed within the hematite grains in the as-sintered sample.

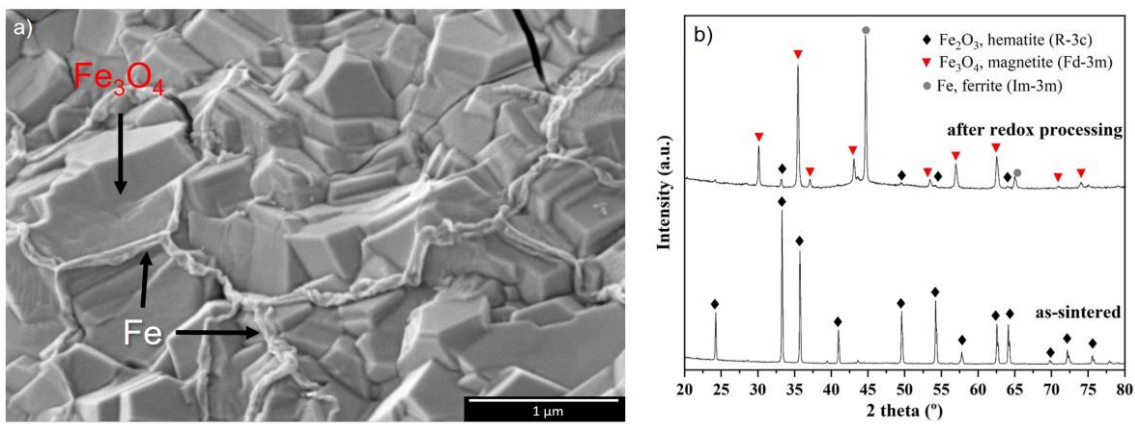

Figure 7. SEM micrograph (a) of the doped $Fe_2O_3$ foams after isothermal redox processing, and XRD patterns (b) for the *as-sintered* and *after redox processing* specimens.

3.5. XPS characterisation.

The XPS spectra shows the valence state and compositional studies of Fe, Al, and Ce elements in the different stages of the doped $Fe_2O_3$ *as-synthesised* (before slurry preparation), *as-fabricated* (before FC and sintering), and *after redox processing* (dynamic). Figure 8 shows the high-resolution photoelectron spectrum for Fe2p (a-c) and Al2p (d). The deconvoluted XPS spectra of the as-synthesised (Figure 8a) and as-fabricated (Figure 8b) doped $Fe_2O_3$ powder shows similar convoluted XPS spectra, where peaks at 709.4, 723.1, and 717.3 eV are identified with the $Fe^{3+}$ ions in α-$Fe_2O_3$ assigned to Fe $2p_{3/2}$, Fe $2p_{1/2}$, and its satellite peak, respectively. The two major peaks for Fe $2p_{3/2}$ and Fe $2p_{1/2}$ are separated by spin energy of 13.7 eV, and the associated satellite peak (717.3 eV) is located at +7.9 eV of the Fe 2p3/2 peak (709.4 eV), which are in good agreement with the reported value for α-$Fe_2O_3$ in the literature [46, 47]. The shake-up satellite (712.8 eV) is also confirmed in the literature for doped $Fe_2O_3$ prepared by the co-precipitation method [48] and with the influence of $Al^{3+}$[49]. In this work, the broadening of the peaks associated with $Fe^{3+}$ to lower binding energies could confirm the change in the oxidation state to $Fe^{2+}$ due to the influence of the $Al^{3+}$ ions being incorporated into the hematite-like $(Fe,Al)_2O_3$ solid matrix. With the sintering process, the satellite peaks of $Fe^{3+}$ in the as-fabricated powder moves to a lower binding energy of 712.5 eV (Figure 8b) due to the local formation of a near-hercynite ($FeAl_2O_4$) phase on the surface of the particles [22].

Subsequent to the dynamic redox process (Figure 8c), the deconvoluted XPS spectrum shows three peaks at 709.5, 716.6, and 723.1 eV corresponding to a non-stoichiometric FeO phase and a peak at 712.1 eV that could be attributed to the aforementioned shake-up satellite. The difference in binding energy between Fe2p 3/2 (709.5 eV) and the satellite peak (716.6 eV) is 7.1 eV, which is 1.1 eV higher than those values reported for oxidation state $Fe^{2+}$ [47, 50]. Nevertheless, the broadening of the Fe2p 3/2 peak might hinder Fe3+ ions corresponding to the $Fe_3O_4$, since the binding energies are similar. In this regard, the XPS spectrum of the *after redox processing* samples shows a significant broadening of the major peak down to lower binding energies, thereby yielding an extra deconvoluted peak at 708.1 eV. This result could be attributed to the presence of metallic iron at the sample surface after the redox cycling [51], which is correlated with the TGA analysis in Figure 4b, and could confirm the spider-like network of Figure 7 as metal Fe. The XPS spectrum of Al2p (Figure 8d) shows a binding energy of 73.8 eV for the as-synthesised samples, although for the as-fabricated and *after redox processing* samples, the binding energy moves to 74.5 eV. In the as-synthesised powder sample, the Al ions could be in a metallic state due to the low temperature of the final heat treatment (800 ºC), which seems sufficient to obtain the $Fe_2O_3$ crystal. However, after having sintered the powder at 1100 ºC and performing the redox processing, the movement of the binding energy to 74.5 eV confirms the formation of $FeAl_2O_4$ on the sample surface [52].

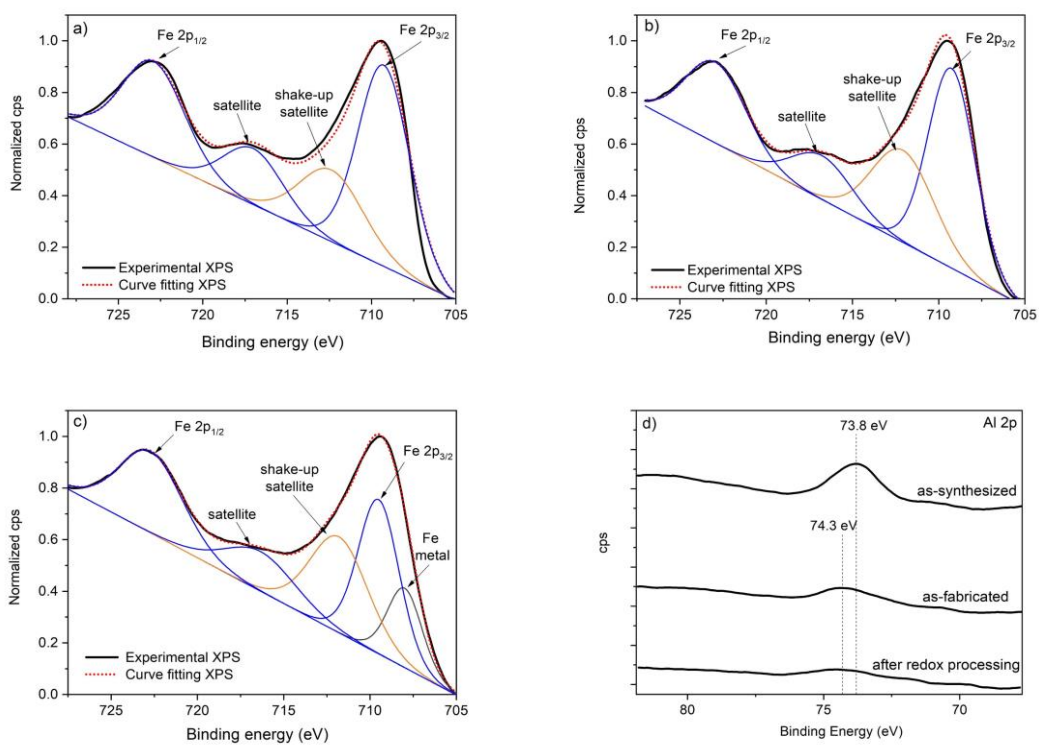

Figure 8. High resolution XPS spectra of Fe2p (a-c) for the doped $Fe_2O_3$ samples as-synthesised (a), as-fabricated (b), and *after redox processing* (c) with the deconvoluted peaks, and Al2p (d) for all samples.

4. Conclusions

Doped Fe2O3 foams with potential redox applications are successfully fabricated by means of FC, and the influences of the Al and Ce atoms in the redox performance are suitably determined.

Al is incorporated into a hematite-like $(Fe,Al)_2O_3$ solid solution in the initial powder, and turns into an $FeAl_2O_4$ phase subsequent to the sintering and redox processing. The Al interaction with iron oxide phases either delays or suppresses the formation of a core-shell structure during sequential redox reaction by reducing the upward mobility of Fe atoms in the cell wall in the doped $Fe_2O_3$ foam during the re-oxidation step in the redox cycling.

No significant influence of $CeO_2$ on the redox performance was established, since Ce atoms were rapidly depleted to the $Fe_2O_3$ particle surface, which limits its interaction during redox cycling.

The pore shrinkage is counterbalance by the development of a highly interconnected pore structure above 100 microns using camphene as the media. By increasing the pore size and narrowing the pore-size distribution, the re-oxidation of Fe to $Fe_3O_4$ is able to accommodate the volume expansion and the exfoliation effects are significantly reduced.

The incorporation of Al atoms into $Fe_2O_3$ by the co-precipitation method seems to more efficiently preserve the pore morphology in the foam subsequent to the redox processing.

Acknowledgements

Financial support for this work has been provided by the Spanish Ministerio de Economía, Industria y Competitividad (MINECO), through the project MAT2016-76713-P. Lloreda-Jurado P.J. also thanks to the Universidad de Sevilla for the financial support (grant PIF II.2A, through VI Plan Propio de Investigación). Furthermore, the authors would especially like to thank Dr. Cristina García-Garrido for her contribution in the XPS analysis and dissertation, and Dr. J. A. Peña & J. Herguido for providing the doped $Fe_2O_3$ powder and their assistance in in the redox assay.